\documentstyle[amssymb,amsmath,12pt,graphicx]{article}

\makeatletter
\@addtoreset{equation}{section}
\makeatother

\begin{document}

\def\text{\mbox}

\begin{center}
{\Huge WAVE ESSENCE OF PARTICLES}

\bigskip \bigskip

D. M. Chi\footnote{%
Present address: 13A Doi Can, Hanoi, Vietnam.}

\emph{Center for MT\&Anh, Hanoi, Vietnam.}

\bigskip \medskip

\textbf{Abstract}
\end{center}

\begin{quotation}
We state several ideas based on the view-point of particle
behaviour of matter to explain wave character of photon and
elementary particles. By using Newton's suggestion of light ray,
we clarify integrally the behaviour of light ``wave''. And
``wave'' character of particles is also explained by the
view-point of particle.
\end{quotation}

\bigskip

\makeatletter
\global\@specialpagefalse
\def\@oddhead{D.M.Chi\hfill Physics/0001036}
\let\@evenhead\@oddhead
% run page numbers, with copyright for first page
\def\@oddfoot{\reset@font\rm\hfill \thepage\hfill
\ifnum\c@page=1
  \llap{\protect\copyright{} 1999 D.M.Chi}%
\fi
} \let\@evenfoot\@oddfoot
\makeatother

\section{Introduction}

Today everybody believes that the matter has both particle
character and wave character.

Following belief up to now, wave is understood as continuous
change of any quantity in period in space and in time with its
immanent cause. With the view-point of particle the periodic
property puzzled everybody, and it is looked as a presence of
wave.

Now, we think that there is still an other way to understand
nature more accurately and more consistently: we use still
Newton's ideas ``along with ray of light there must be a
manifestation of some periodicity'' to explain wave phenomena of
the light and elementary particles.

The article is organized as follows. In Section 2, we show an
explanation intuitionally of interference of the light. And the
interferential phenomenon of electron is illustrated in Section 3.
Conclusion is given in Section 4.

\section{Interferential picture of the light}

The electromagnetic field can exit independently and so it
includes invariant structures (particles). The electromagnetic
field has periodicity and so this periodicity either goes with
particle by anyway or manifests in the distribution of particles
in space and this squadron of particles fly along a fixed ray,

\begin{center}
\begin{tabular}{ll}
$|$ --- $\lambda $ -- $|$ & $\overrightarrow{c}$
\\
$\bullet$-- -- -- -- $\bullet$-- -- -- -- $\bullet$-- -- -- --
$\bullet$ & -- -- -- $\rightarrow $%
\end{tabular}
\end{center}

This picture is similar to the so-called wave.

Let us use this imagination of the light to explain how the
interferential picture is created.

Suppose that there is a gun, and after each fixed interval of time
$\Delta t$ it shoots one ball. Balls are alike in all their
aspect. They fly with the same velocity $v$ in environment without
resistance force and their gravities are ignorable.

The ``wave-length'' - i.e. the distance between two balls - is
$\lambda =v.\Delta t$. All balls are electrized weak charge of the
same sign, and they are covered with a sticky glue envelope so
that electrostatically propulsive force between them cannot win
stickiness of glue envelope when they touch together with small
sufficient meeting angle. A target with two parallel interstices
in the vertical is set square with the axis of the gun and is also
electrized weak charge but different from the sign of balls. The
distance between two interstices is not too large in comparison
with the wave-length $\lambda $. And assume that sticky glue of
balls does not effect on the target.

Thus, probability in order that any ball flies through one
interstice or the other is identical. (The interstices are enough
large that balls can fly through easily).

After flying past, balls are changed flying direction with all
possible angles in the horizontal plane with definite probability
distribution.

With such initial conditions, balls are in `dephasing' with each
other and, after overcoming the target, balls that do not fly
through the same interstice are able to meet together with a some
non-zero propability. If meeting angle is enough small in order
that stickiness has effect, two some balls will couple together to
be a system, then change direction and fly on the bisector of
meeting angle.

If behind the target and distant from the target a space $d\gg
\lambda $, a reception screen is set parallel to the target, then
on this screen we will harvest falling point locations of single
balls and couple balls.

Argumentation and calculation show that single balls (missed
interference) form a monotonous background of falling points, and
couple balls (caused by interference) fall concentratively and
create definite veins on the background, depending on dephasing
degree of interfered balls.

Hence, from Newton's idea and using quantities such as
``wave-length'', ``dephase'', and so on results obtained is fitted
in ones calculated using wave behaviour. Furthermore, they explain
why, when amount of balls is not enough large (the time to do
experiment is short), the picture of falling points seems chaotic,
randum. Only with a large number of balls (the time to do
experiment is long), interferential veins are really clear. This
is one that, if using wave behaviour, is impossible to explain.

Thus, if we consider the light as a system of particles that, in
the most rudimentary level, are similar above balls: they are
attractable together and radiated periodically, then the
interference of the light is nothing groundless or difficult to
understand when we refuse to explain it by using wave behaviour.
Moreover, without wave conception the phenomena of the light is
very bright, clear, and more unitary.

Such a view-point of the light requires to imagine again many
problems, simultaneously brings about new effects that need to
prove in experiment.

If there is a source that radiates continously separately light
pulses of a fixed thickness and with a fixed distance between
pulses (Fig. 1), we can carry out a following experiment (Fig. 2).

\begin{figure}[h]
\begin{center}
\leavevmode
\includegraphics[width=0.2\columnwidth]{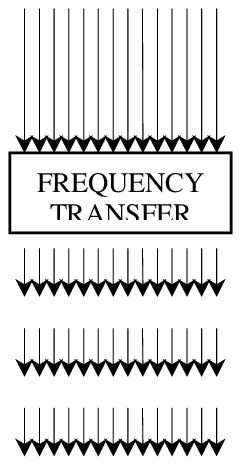}
\caption{Frequency transfer.}
\end{center}
\end{figure}

With two continously radiative sources we direct light pulses
together with an intersecting angle $\alpha $. Interference
presents only in the area ABCD of Fig. 1. If the light is wave,
then to see interference we should set a photographic plate in the
area ABCD, because outside this area interference is impossible to
present. Two light sources are independent from each other, the
stable condition of interferential veins is not ensured, position
of vains is changed incessantly, and the consequence is that on
the photographic plate we cannot obtain interferential veins.

\begin{figure}[h]
\begin{center}
\leavevmode
\includegraphics[width=0.5\columnwidth]{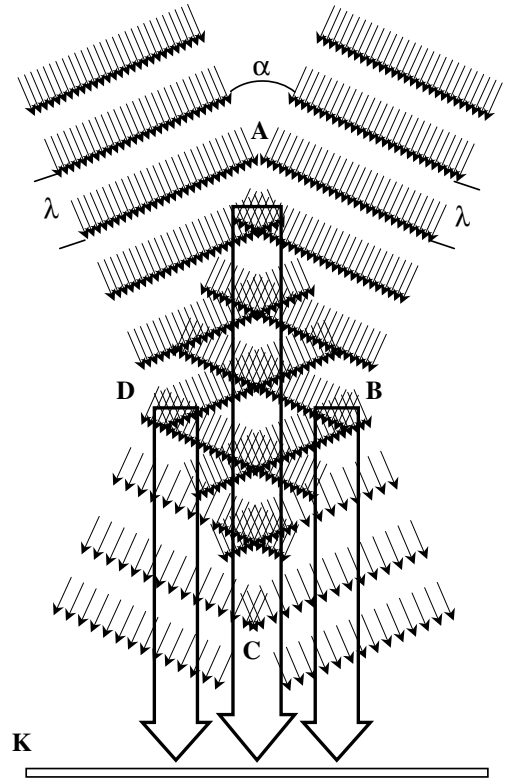}
\caption{Interference of the light quanta.}
\end{center}
\end{figure}

But if the light is particle, then we always obtain interferential
veins (in suitable polarization condition) though the photographic
plate is set inside or outside the area ABCD.

Carrying out experiments according to this diagram we can check
interferential ability of the light of different wave-lengths.
Given two radiative sources of different frequencies by that way,
we are able to see whether photons of different frequencies is
identical.

A consequence of particle behaviour is that we can make
mirror-holography with any reappear light source. This is drawn
from the phenomenon that two photons interfered together change
direction and fly on the the bisector of meeting angle.

In the conception of particle, frequency of a light wave is
understood as number of photons radiated in a unit of time to a
definite direction. The photo-electric effect, therefore, is
understood as follows: the more number of coming photons that
collide to electrons in a unit of time is, the higher energy that
electrons gain is. If momenta of all photons are of the same
value, then it is possible that number of photons electrons gain
is proportional to light frequency, and thus momentum of each
photon is proportional to Plank's constant $\hbar $. With such
understanding, energy that electron gain from interfered photons
is higher than one from non-interferential photons.

With wave behaviour and imagined that atom sends spherical waves,
then the photo-electric effect is impossible to explain as shown.
But if thought that atom radiate directly light, then because
radiation is a wave process, in radiation process atom does not
keep still a place in space, thus it is difficult to maintain that
all energy quanta is transmitted out space in a definite ray.
Situation is the same for absorbing process of energy quanta. For
instance, electron is impossible to keep still a place to await
absorbing all energy quanta then moves to other position.

Thus, there is not any firm basis to say that energy is absorbed
piecemeal quanta $E=\hbar \nu $.

\section{Interferential picture of elementary particle}

Let us consider interference of electron.

First of all we can confirm that there would not be any experiment
we gain interference of electron if we used conditions as already
stated for the light. Because with actual experience electrons are
not like as above balls with stickiness that lack of this ability
there is not any presence of interferential couples.

Interference of electron is completely different, if using the
word ``interference''.

Suppose that there is an electron flying to a block of matter made
up from particles heavier very much than electron. Each heavy
particle in the block of matter is a scattering center. Because of
interaction, after flying out of influential region of scattering
center electron is changed direction with an angle $\alpha $ in
comparison with initial direction.

Assume that the deviation angle $\alpha $ is dependent on the aim
distance $\rho $ obeying on the law \textit{a} or \textit{b} as on
the figures.

\begin{figure}[h]
\begin{center}
\leavevmode
\includegraphics[width=0.5\columnwidth]{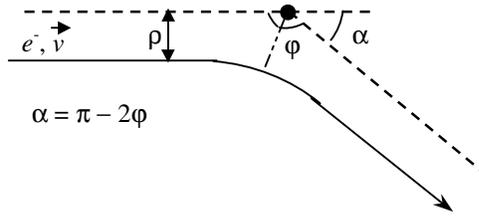}
\caption{Scattering of electron.}
\end{center}
\end{figure}

\begin{figure}[h]
\begin{center}
\leavevmode
\includegraphics[width=0.7\columnwidth]{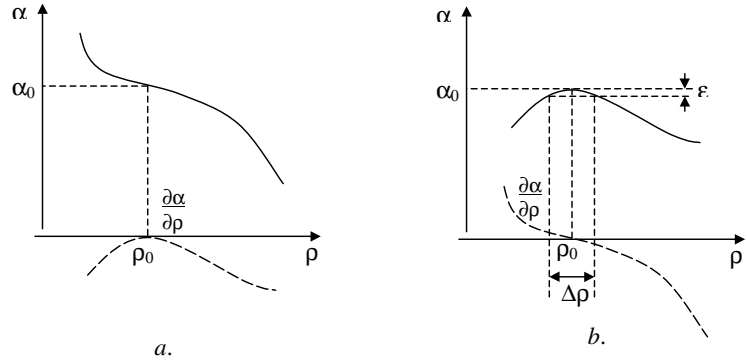}
\caption{$\alpha$-dependence of the aim distance $\rho$.}
\end{center}
\end{figure}

Derivative of $\alpha $ with respect to $\rho $ is equal to $0$ at
$\rho =\rho _{0}$.

If all values of the aim distance $\rho $ is the same probability
for $e$, then the probability in order that the particle $e$ is
deviated from the coming direction an angle ($\alpha _{0}$) is
``infinitely large'' in comparison with any other angle: $\left.
\frac{\partial \rho }{\partial \alpha }\right| _{\alpha =\alpha
_{0}}=\infty $.

It is necessary to say that $\alpha _{0}$ is inversely
proportional to the momentum of $e$. The higher the momentum of
$e$ is, the higher the direction conservability of its momentum
vector is, then the smaller the deviation angle $\alpha $ is.
(That is just the basis of Broblie relation.)

We bring in a quantity $\epsilon $ called the maximum divergence
that measurement is acceptable, that border rays deviated in
comparison with the angle $\alpha $ a value $\epsilon $ belongs to
still that angle $\alpha $. Thus, we can establish a function of
probability density having the following form, (Fig. 5),

\begin{figure}[h]
\begin{center}
\leavevmode
\includegraphics[width=0.4\columnwidth]{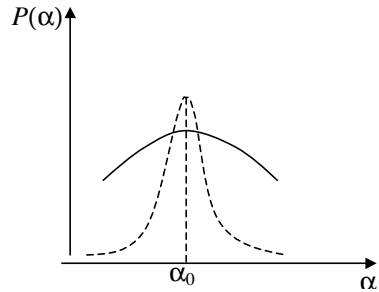}
\caption{Probability density.}
\end{center}
\end{figure}

The greater the sharpness of this distribution is, the greater the
contrast of probability densities between the angle $\alpha $ and
its neighbouring angles.

Call average distance between two scattering centers $2R$, then
influence space of any center (in the field of the aim distance)
is $\pi R^{2}$.

The space, where the particle $e$ is falled into and deviated an
angle ($ \alpha \pm \epsilon $), is proportional to $\pi \left[
(\rho +\Delta \rho )^{2}-(\rho -\Delta \rho )^{2}\right] \approx
4\pi \rho \Delta \rho $.

\begin{figure}[h]
\begin{center}
\leavevmode
\includegraphics[width=0.25\columnwidth]{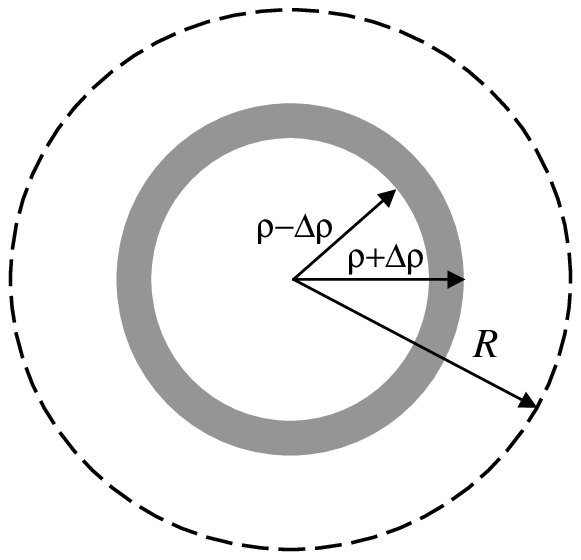}
\caption{}
\end{center}
\end{figure}

The probability in order that $e$ scatters into the angle $\alpha
$ is $ \frac{4\pi \rho \Delta \rho }{\pi R^{2}}=\frac{4\rho \Delta
\rho }{R^{2}}$ (Fig. 6).

The probability density that $e$ scatters into the angle $\alpha $
in a some direction is $\frac{4\rho \Delta \rho
}{R^{2}}\frac{1}{2\pi }=\frac{4\rho \Delta \rho }{\pi
R^{2}}=P_{(\alpha )}$.

However, using the above method to estimate the probability
density with all directions in the space for all scattering
processes of particles is very completed and unwieldy. For this
reason, here we are only interested ``relative'' probabilities of
variety directions, namely events: the probability in order that
$e$ scatters into the angle $\alpha _{0}$ is
``infinitely''\footnote{%
It means $\epsilon \rightarrow 0$.} large in comparison with that
for all other angles. In that correlation, all deviation angles
that are different from $\alpha _{0}$ should be ignored because
they form only a monotonous background. This has not influence on
the qualitative accuracy of the law.

Consider $i$-th scattering experiment (each experiment there is
only one $e$ attended with a constant momentum, the same for all
experiments.)

Assume that all scatterings are elastic. The scattering process of
$e$ is easy to imagine as Figures 7 and 8.

\begin{figure}[h]
\begin{center}
\leavevmode
\includegraphics[width=0.7\columnwidth]{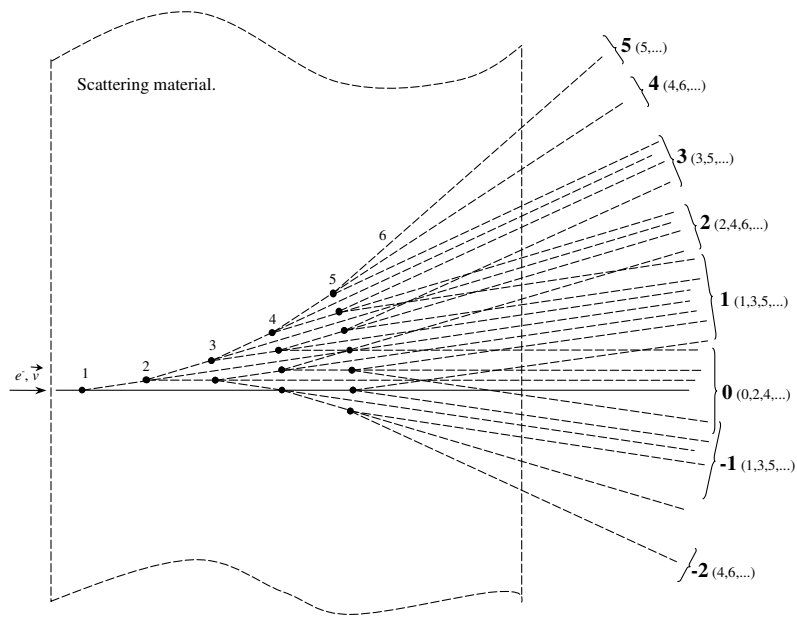}
\caption{Projection outline of highest-probability diffraction
beam on the vertical space.}
\end{center}
\end{figure}
\begin{figure}[h]
\begin{center}
\leavevmode
\includegraphics[width=0.7\columnwidth]{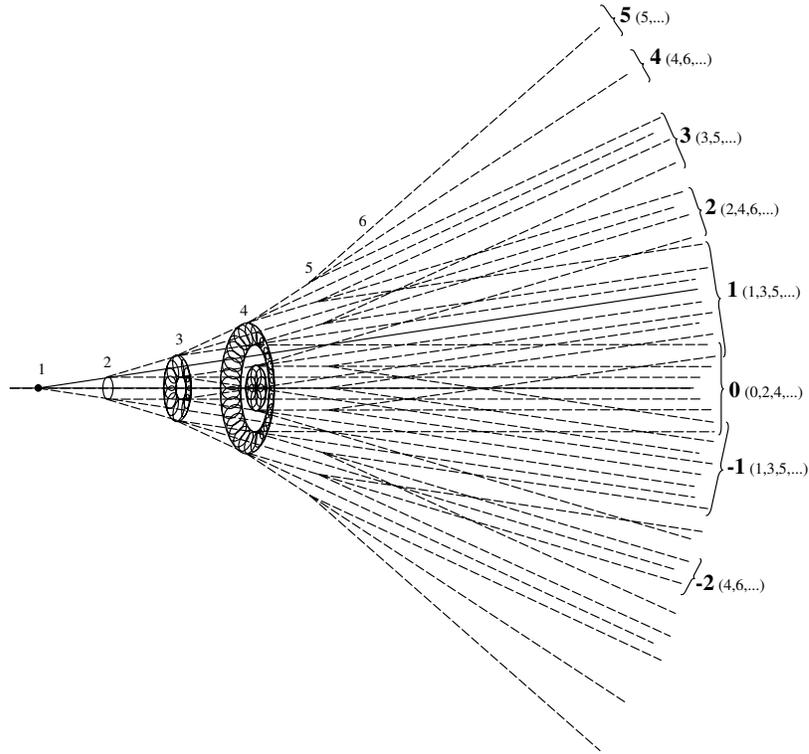}
\caption{Reproduction of the formation of electron's diffraction
veins.}
\end{center}
\end{figure}

After the 1-th scattering, $e$ can be deviated with any angle and
fly with any direction, but all directions with highest
probability ($P_{\alpha _{0}}=\infty $) form a cone with the top
at the scattering center and an arrangement angle ($2\alpha
_{0}$).

In the 2-nd scattering, the directions with highest probability of
$e$'s trajectory form the cones with angles ($0$, $2(2\alpha
_{0})$). This is explained as follows.

Scattering centers in the block of matter distribute at random for
$e$'s trajectory and this random is always maintained by thermic
fluctuations, inelastic scatterings, ... So, on the conic surface
(1-2) there forms a brim - the locus of probability of 2-nd
scattering centers, is plotted as the brim (2-2) in the figure. At
each point of the brim there exist 2-nd scattering centers with
some probability.

Thus, in the 2-nd scattering at each point of the brim there forms
a new probability cone, similar to fomation at the 1-th scattering
center. These probability cones interfere with each other forming
two collective cones with open angles $0$ and $2(2\alpha _{0})$
respectively.

Indeed, if we put a spherical surface of enough large radius and
center coincident with the 1-th scattering center, then cones
intersect the spherical surface and form circles as given in
Figure 9.

\begin{figure}[h]
\begin{center}
\leavevmode
\includegraphics[width=0.5\columnwidth]{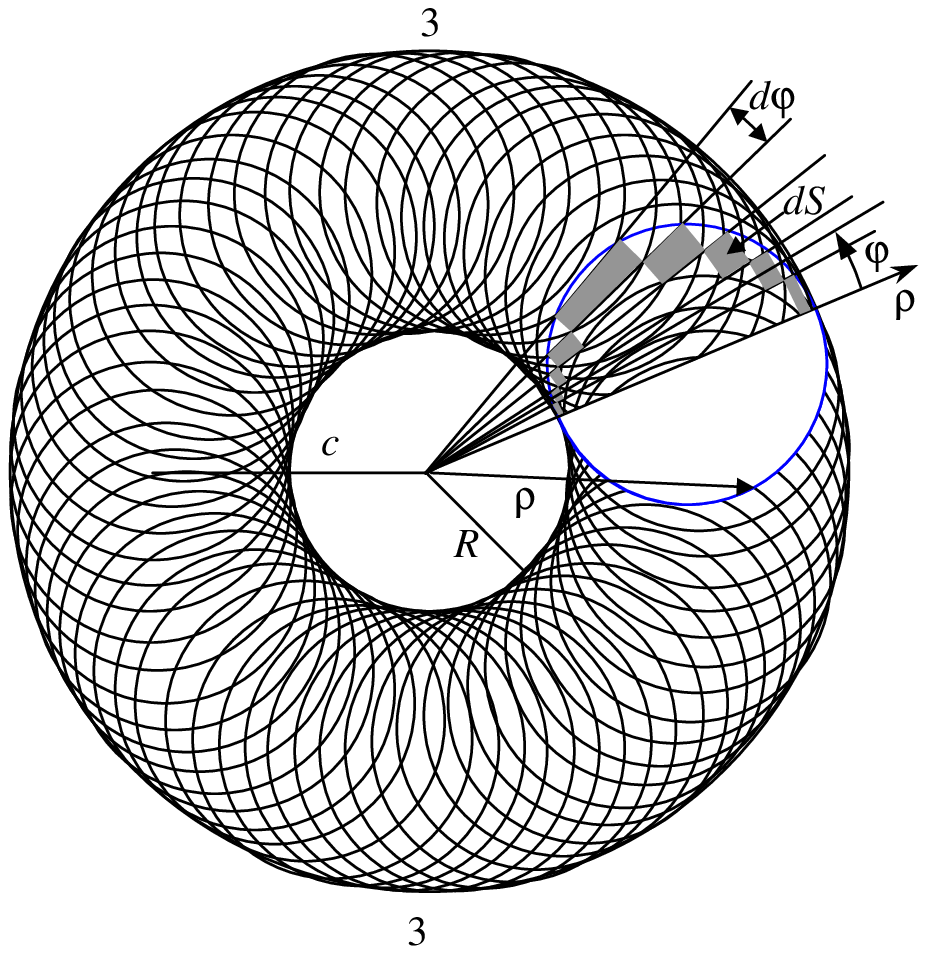}
\caption{}
\end{center}
\end{figure}

For simplification, we are not concerned with curvature of the
spherical surface. The probability in order that the particle
falls into any point of circles is identical.

Let us find the probability density of $e$'s falling points on
this surface. In the figure, it is clear that this density is
proportional to the ratio $ \frac{\Delta \ell }{\Delta S}$, where
$\Delta \ell $ is the total length of circles in an elementary
surface $\Delta S$ ($dS$).

Solve this problem in the pole coordinate, the equation of circles
in this coordinate is
\begin{eqnarray}
\rho &=&C\cos \varphi \pm \sqrt{R^{2}-C^{2}\sin ^{2}\varphi }, \\
d\rho &=&-C\sin \varphi \left( 1\pm \frac{C\cos \varphi }{\sqrt{
R^{2}-C^{2}\sin ^{2}\varphi }}\right) d\varphi , \\ dS &\cong
&\rho d\rho d\varphi ,  \notag \\ \Delta \ell &\cong &2\sqrt{d\rho
^{2}+\rho ^{2}d\varphi },  \notag \\ M_{\rho} &=&\frac{\Delta \ell
}{\Delta S}=\frac{2\sqrt{d\rho ^{2}+\rho ^{2}d\varphi }}{\rho
d\rho d\varphi }.
\end{eqnarray}

By differentiating of $M_{\rho}$ over $\rho $ then putting this
derivative zero or substituting values of $\rho $ and $d\rho $
into (3) and approach $\varphi \rightarrow 0$ we also obtain $
M_{\rho}\rightarrow \infty $.

Thus, at maximum and minimum values of $\rho $ the probability
density $ M_{\rho} $ is infinitely large (in comparison with other
values of $\rho $).

The surface $dS$ is equivalent to the volume angle $d\Omega $, and
$\Delta \ell $ is equivalent to the probability in order that the
particle scatters into that volume angle.

Therefore, we obtain that in the 2-nd scattering all possible
directions with highest probability of trajectories form two cones
with open angles $0$ and $2(2\alpha _{0})$.

With similar argument we realize that at the 3-rd scattering the
directions with highest probability of $e$'s trajectories form
cones with open angles $ 3(2\alpha _{0})$, $1(2\alpha _{0})$,
$1(2\alpha _{0})$, $-1(2\alpha _{0})$.

Generally, up to the $n$-th scattering there form probability
cones of possible directions of $e$'s trajectories with open
angles $n(2\alpha _{0})$ , $(n-2)(2\alpha _{0})$, ...,
$(n-2m)(2\alpha _{0})$.

Put a spherical surface as a catching screen with its axis
coincident to the initial direction of particle before coming to
the target, and its radius much larger than the thickness of the
target. The center of the spherical surface is on the target and
in the coming point of scattering particle.

Hence, the probability cones forming at the last scattering
intersect the catching screen and form circle brims - These are
locus of falling points with $e$'s maximum probability.

If $n$ is even, we obtain an even number of brims; $n$ is odd, we
have an odd number of brims. And in all times of experiment $e$
always scatters $n$ times, then on the catching screen we obtain
either an even number or odd number of brims, depending on even or
odd value of $n$. Of course, in one time of experiment, $n$ has
only value. Thus, the probability intensity of obtained brims
after total experiment $\Sigma $ is dependent not only on
decreasing law from inner to outer but also on frequence of $n$,
i.e. on the ratio $\frac{i_{n}}{\Sigma }$; $i_{n}$ is frequency
(number of occuring times of $n$) in total experiment $\Sigma $.

It is difficult to define this frequency for every possible value
of $n$. However, at the most rudimentary, it is sure that spectrum
of $n$'s values is not large\footnote{ We are only interested
particles that overcome through the target.} and probability in
order that $n$ has even or odd value is identical, then spectrum
of differaction brims is complete from ($0$) to
($n_{\text{max}}$).

From conditions of experiment we realize that the larger the
matter density of the target is the tickness is, the larger
$n_{\text{max}}$ is.

Here, once again we find out that the differaction picture is only
clear since the number of particles taking in scattering are
enough much. If they are is too little, on the screen we see only
a chaotic distribution of $e$'s marks, but on the contrary, if
they are too much and the catching screen is a photographic plate,
then all points on the screen are saturated and differaction brims
are hidden.

The scattering of light particles in a radial field is described
in mechanics as follows
\begin{equation*}
\varphi
=\int_{r_{0}}^{r}\frac{M}{mr^{2}}\frac{dr}{\sqrt{\frac{2}{m}\left(
E- \frac{M^{2}}{2mr^{2}}-U(r)\right) }},
\end{equation*}
where $\varphi $ is the angle made by radius vector of particle's
position on the trajectory and radius vector of particle's
extremal point ($\vec{r} _{0}$), $M$ is the momentum of the
particle, $E$ is the energy and $m$ is mass of the particle, and
$U(r)$ is the potential of the field.

If the potential of the field $U(r)$ has the form $U(r)=\pm
\frac{A}{r}$, then from the above formula we can express $\varphi
$ as a function of the aim distance $\rho $ since the upper bound
approaches to infinity, thus the deviation angle $\alpha =\pi
-2\varphi $ is also a function of $\rho $.

Calculations give the result that the derivative of $\varphi (\rho
)$ with respect to $\rho $ is not equal to zero at any position.
That proves that the condition to have ``wave-like'' differaction
is not satisfied.

If in fact the potential of nuclear field had the form as above,
then the ability in order that electron would fall into nucleus
has a very large probability. This is not compatible with the
fact\footnote{ To avoid this there was a quantum mechanics.}.

We can suppose a supplement (not interested in quantum mechanics)
as: far from the attracting center a distance $a$ there is a
surface $L$. This surface changes trajectory of scattering
particles. Because it is not absolutely hard (in the present
region of the surface, the potential field has some form), there
happens a slippery effect of particle on the surface. This softens
variation of deviation angle of trajectory $\alpha $, and then
$\alpha $ is still a continous function of $ \rho $. Thus, the
momentum of scattering particle is unsurpassed a some value in
order to unbreak elasticity of the surface. Otherwise, the
momentum of scattering particle is larger than a supposed
criterion, the surface $L$ is broken to form the light radiation.
Some radiation forms can belong to Trerenkov or Compton effect.

In summary, with this supplement, $\alpha $ is a function of $\rho
$ with dependence as Figure 10.

\begin{figure}[h]
\begin{center}
\leavevmode
\includegraphics[width=0.25\columnwidth]{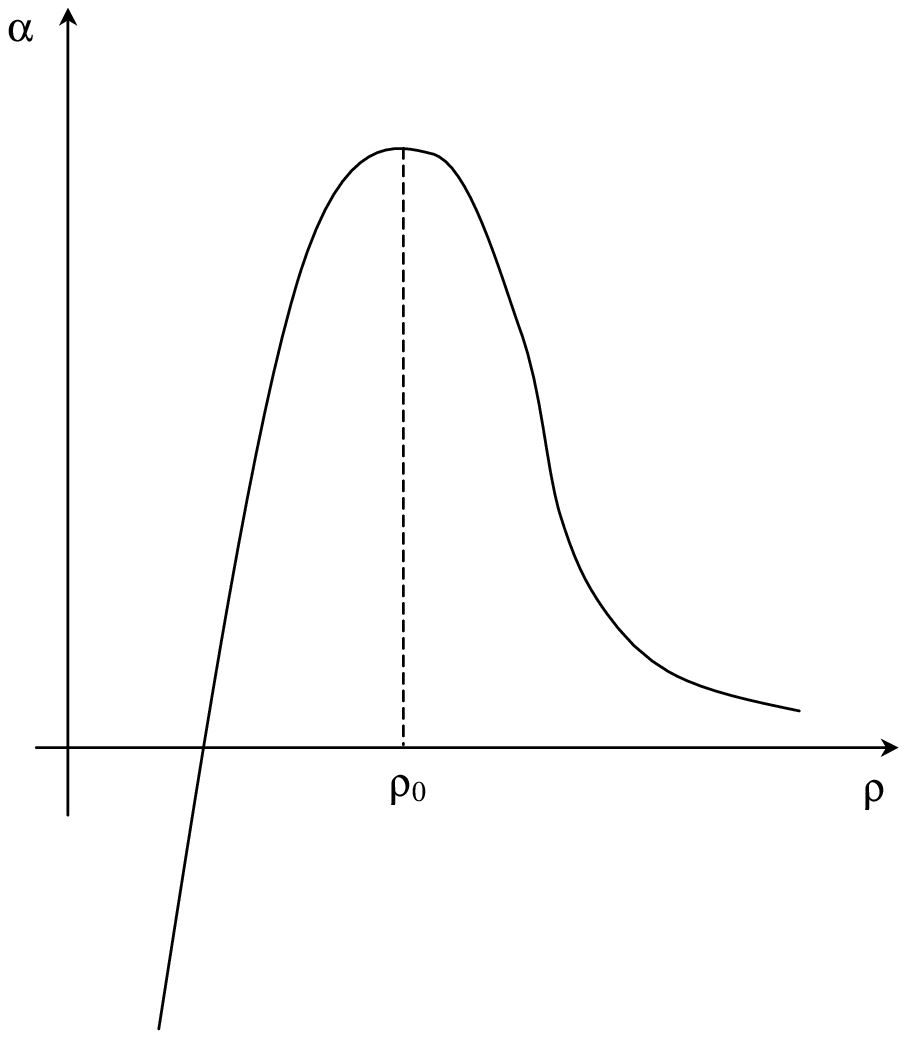}
\caption{}
\end{center}
\end{figure}
where $\rho _{0} \approx a$ and hence, the condition to have the
``wave-like'' differaction is satisfied.

\section{Conclusion}

Thus, we show in rather detail some ideas based on the particle
behavour of matter. By using Newton's model of light ray we have
explained rather completely ``wave'' behaviours of the light. The
polarization of the light is a effect of particle behaviour: two
photons interfere with each other forming a system with
axisymmetry. The experiment set as Figure 1 is important. Carrying
on this experiment will take part in confirmation of light's
particle behaviour. Its detailed results will open up new ideas,
and new directions of research.

The wave character of elementary particle, electron is typical,
can be explained by pure particle behaviour. This give us a
similarity between the wave function in quantum mechanics and the
vector function of particle behaviour.

The motion of any particle can be expressed as a vector, whose
direction points out particle's motion direction at a given point
of trajectory, and whose module expresses probability amplitude of
particle flying in that direction. Thus, the probability
trajectory of particle is completely able to expressed as a vector
function
\begin{equation*}
\psi =A(\alpha _{(t)})e^{i\alpha _{(t)}},
\end{equation*}
$\alpha _{(t)}$ is the deviation angle in comparison with the
initial direction, is a function of time; $A(\alpha )$ is the
probability of particle flying with the angle $\alpha $.

But this does not mean that the particle expressed as above will
have a really wave character.

If we find a way to express the value of $\alpha _{(t)}$ by the
value of energy-momentum of scattering particle, parameters of
scattering environment, and simplification: $\alpha _{(t)}$ is
continously independent of time but discontinously dependent on
time (due to the fact that scattering centers are discontinous),
then the vector function is not basically different from the wave
function in quantum mechanics. Doing with appropriate operators
for probability function, we can obtain correlative quantities.

Hence, from the natural idea of particle behaviour of matter we
can discover further natural phenomena. One of the most host
problems today is inflation of the universe affirmed from the
red-shift of Doppler's effect. However, if the space between
observer and light source is vacuum, then the explanation of the
red-shift based on Doppler's effect is fully satisfactory. But in
fact the universe is filled with gravitational fields, macrometric
and micrometric objects as stars and clusters. Therefore, these
problems had been re-examined in further detail by us, with
consideration actual influences of interstellar environment on
frequency shift. This is only realized by foundation of particle
behaviour of the light. The existence of cosmic dusts as motional
scattering centers is an essential condition to happen
scattering-interference processes of the light when it flies
through the interstellar environment. In turn, these
scattering-interference processes lead to the shift of light
frequency. Our results give a reliable confirmation that the
red-shift is not exhibit of the inflation of the universe.

\section*{Acknowledgments}

The present article was supported in part by the Advanced Research
Project on Natural Sciences of the MT\&A Center.

\end{document}